\documentclass[]{iopart}
\expandafter\let\csname equation*\endcsname\relax
\expandafter\let\csname endequation*\endcsname\relax
\usepackage{amsmath}
\usepackage{iopams}
\usepackage{graphicx}
\usepackage{color}
\usepackage[english]{babel}
\usepackage{citesort}
\usepackage{cite}
\usepackage{ulem}
\usepackage{upgreek}
\usepackage{soul}
\usepackage{siunitx}

\usepackage[colorlinks=true,linkcolor=blue,anchorcolor=blue,citecolor=red,filecolor=green,menucolor=blue,runcolor=blue,urlcolor=blue]{hyperref}

\usepackage[utf8]{inputenc}
\usepackage{graphicx}
\usepackage[colorinlistoftodos]{todonotes}

\usepackage{notes2bib}
\usepackage{tikz}
\renewcommand{\epsilon}{\varepsilon}

\begin{document}
\title[A digital four-arm bridge  for the comparison of resistance with capacitance] {A digital four-arm bridge for the comparison of resistance with capacitance}
\author{Yicheng Wang and Stephan Schlamminger}
\date{April 2024}
\address{National Institute of Standards and Technology,
100 Bureau Drive, Gaithersburg, MD, USA}
\ead{ywang@nist.gov, stephan.schlamminger@nist.gov}
\begin{abstract}
We have built and demonstrated a digital four-arm bridge for the comparison of resistance with capacitance. The digital four-arm bridge mimics the classical quad bridge in the digital domain with three balances: the source balance, the detector balance, as well as the main balance. Due to correlation, the required precision of the source voltages is only of the order of the square root of the ultimate bridge precision. For the comparison of a \SI{100}{\kilo\ohm} resistor with a \SI{1}{\nano\farad} capacitor near \SI{1592}{\hertz}, the combined standard uncertainty $(k = 1)$ is \SI{5e-9}{}.  
\end{abstract}
\maketitle
\ioptwocol

\section{Introduction}

A recent review article, entitled ``Impedance bridges: from Wheatstone
to Josephson'', summarized the state of the art of impedance bridges~\cite{Overney2018}.  However, one bridge did not receive the recognition it deserves. A six-arm bridge invented by Greig Small {\it et al.} for the comparison of resistance with capacitance ($R-C$) was published in this journal more than 20 years ago~\cite{Small2001}. The main bridge loop consists of two $R-C$ arms, two $C-R$ arms, and two arms of a ratio transformer which is one-to-one with respect to its center tap through which the in-phase and quadrature adjustment voltages are injected. The bridge has several salient features. It does not need any defining transformers that have been widely believed to be essential for accurate four-terminal-pair (4TP) impedance measurements~\cite{Cutkosky1970,Awan2011}. It also defines a common low-potential port between the $R-C$ arms and another between the $C-R$ arms, thereby saving the combining networks between the low-current ports and the low-potential ports that are commonly used to eliminate the effect of contact resistance and connecting cables. Despite these simplifying measures, the bridge reached extremely low standard uncertainties of better than 1 part in $10^9$ at frequencies from \SI{200}{\hertz} to \SI{2}{\kilo\hertz}.

Different from the conventional passive-component quad bridge~\cite{Cutkosky1970,Thompson1968}, the six-arm bridge uses active components to synchronously generate three excitation sources that minimize harmonics, and it also uses active components to produce an appropriately weighted sum of three detector currents that reflect the bridge balance error. It retains the most important feature of the conventional quad bridge with three balances: the main bridge balance, the source balance, and the detector balance, where the precision required in setting the source voltages and the detector combining network parameters is only of the order of the square root of the ultimate bridge precision. This weak dependence of the bridge performance on the precision of the voltage sources is in stark contrast with the digital bridges where the voltage ratio of two synthesized sources is used directly as the reference for impedance ratio measurements and its stability can become a major limiting factor for the overall bridge performance~\cite{Overney2016,Marzano2022}.

In this paper, we present a digital four-arm bridge for comparing resistance with capacitance. The source balance and the detector balance are implemented using digital techniques. 
We demonstrate that these auxiliary balances contribute only second-order errors to the main bridge balance.

\section{Theory of the four-arm bridge}

In 1929, Ogawa~\cite{Ogawa1929} combined the Wagner balance~\cite{Wagner1911} with Carvallo's theorem~\cite{Carvallo1925} and theoretically proposed the fully balanced four-arm bridge. The significant conclusion of Ogawa's work is that the bridge resolution is a product of two small terms, the source balance error and the detector balance error. This was experimentally demonstrated in 1940 by Koops~\cite{Koops1940}. At the fourth CPEM, held in Boulder, CO, in 1964, Thompson~\cite{Thompson1964} demonstrated a quad bridge, which is a variation of the four-arm bridge and shares the same weak dependence on the source and detector balances. With the advent of precision ratio transformers in the 1950s, the research focus shifted away from the traditional four-arm bridge. Here, we revive the classical four-arm bridge with digital techniques.

\begin{figure}[htbp]
\begin{center}
\includegraphics[width=0.95\columnwidth]{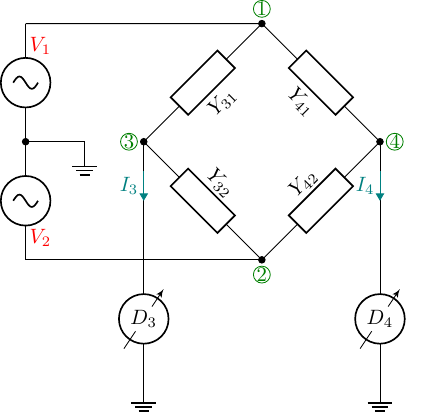}
\end{center}
\caption{ Schematic of a coaxial four-arm bridge where the outer conductor has been omitted for clarity. The  current detectors $D_3$ and $D_4$ force node 3 and node 4 to virtual ground.
\label{fig:simple4}}
\end{figure}

For a general coaxial four-arm bridge illustrated in Fig.~\ref{fig:simple4}, the passive bridge network is a four-port system that can be completely described by four linear equations, including the effect of ground admittances at the four nodes~\cite{Ogawa1929}. We consider a special case where the detector inputs, port 3 and port 4, are virtual-ground current inputs to avoid dependence of the bridge network on ground admittances at the detector nodes. The bridge can then be described by a system of two linear equations,
\begin{eqnarray}
I_3 &=& Y_{31} V_1 + Y_{32} V_2 \label{eq1}\\
I_4 &=& Y_{41} V_1 + Y_{42} V_2, \label{eq2}
\end{eqnarray}
which  can  be  written more compactly in matrix form:
\begin{equation}
\mathbf{I} =\mathbf{Y} \mathbf{V} \label{eq:I}.
\end{equation}

Following~\cite{Small2001}, we can combine $I_3$ and $I_4$ to generate an error current, $\Delta I$, which reflects deviations from the perfect bridge balance,  i.e., $\Delta I=0$. 
The error current is a linear combination with complex coefficients of the two detector currents, given by
\begin{equation}
\Delta I =
\left(
\begin{array}{cc}
\rho_1 & \rho_2
\end{array}
\right)
\left(
\begin{array}{c}
I_3 \\ I_4
\end{array}
\right) = 
\boldsymbol\rho \mathbf{I} \label{eq:eps}.
\end{equation}
The error current is a virtual quantity, meaning it only exists mathematically in the measurement computer and is not obtained by physically combining currents. as was done in~\cite{Small2001}.

Combining Eq.~(\ref{eq:I}) and  Eq.~(\ref{eq:eps}) yields
\begin{equation}
\Delta I=\boldsymbol\rho\mathbf{Y} \mathbf{V}. \label{eq:vareps}
\end{equation}
The error current $\Delta I$ is zero and is independent of $\mathbf{V}$ if 
\begin{equation}
\boldsymbol\rho\mathbf{Y}  =\mathbf{0}, \label{eq:db}
\end{equation}
which has been called the detector balance~\cite{Thompson1968}. Similarly, $\Delta I=0$ and is independent of $\boldsymbol\rho$ if 
\begin{equation}
\mathbf{Y}\mathbf{V}  =\mathbf{0}, \label{eq:sb}
\end{equation}
which has been called the source balance. For nontrivial solutions to Eq.~(\ref{eq:db}) and Eq.~(\ref{eq:sb}), $\mathbf{Y}$ must be singular, i.e.,
\begin{equation}
\det{(\mathbf{Y})} =0 \label{eq:mb}
\end{equation}
which is the main bridge balance. 

Let $\mathbf{Y}= \mathbf{Y}_\circ$  be a solution to  Eq.~(\ref{eq:mb}), $\boldsymbol\rho= \boldsymbol\rho_\circ$  be  a solution to $\boldsymbol\rho\mathbf{Y}_\circ  =\mathbf{0}$, and $\mathbf{V}= \mathbf{V}_\circ$  be a solution to $\mathbf{Y}_\circ\mathbf{V}  =\mathbf{0}$. Then, with $\mathbf{Y}=\mathbf{Y}_\circ + \boldsymbol\delta\mathbf{Y}$, $\boldsymbol \rho=\boldsymbol \rho_\circ + \boldsymbol\delta\boldsymbol \rho$, and 
$\mathbf{V}=\mathbf{V}_\circ + \boldsymbol\delta\mathbf{V}$, Eq.~(\ref{eq:vareps}) becomes
\begin{eqnarray}
\Delta I &=&
\left(\boldsymbol\rho_\circ + \boldsymbol\delta\boldsymbol \rho\right)
\left(\mathbf{Y}_\circ + \boldsymbol\delta\mathbf{Y}\right)
\left(\mathbf{V}_\circ + \boldsymbol\delta\mathbf{V}\right)\nonumber\\
&=&
\boldsymbol\rho_\circ \boldsymbol\delta\mathbf{Y} 
\mathbf{V}_\circ\nonumber\\
&&+ 
\boldsymbol\rho_\circ 
\boldsymbol\delta\mathbf{Y} 
\boldsymbol\delta\mathbf{V} 
+
\boldsymbol\delta\boldsymbol \rho
\mathbf{Y}_\circ
\boldsymbol\delta\mathbf{V} 
+
\boldsymbol\delta\boldsymbol \rho
\boldsymbol\delta\mathbf{Y} 
\mathbf{V}_\circ
\nonumber\\
&&
+
\boldsymbol\delta\boldsymbol \rho
\boldsymbol\delta\mathbf{Y} 
\boldsymbol\delta\mathbf{V}.
\label{eq:epsfull}
\end{eqnarray}
Importantly, $\boldsymbol\delta\mathbf{Y}$ contributes to the first-order, while $\boldsymbol\delta\mathbf{V}$ and $\boldsymbol\delta\boldsymbol\rho$ contribute to second and higher order terms of $\Delta I$. Thus, the bridge is sensitive to the impedance ratio to first order but is less sensitive to the source voltage fluctuation and the linear combinations of the currents.

According to Eq.~(\ref{eq:mb}), if $Y_{31}$ and $Y_{42}$ are capacitors with the values $C_1$  and $C_2$, respectively, and $Y_{32}$ a resistor with the value $R_1$ and the bridge is operated at $\omega$,  one finds
\begin{eqnarray}
\det{(
\mathbf{Y}_\circ)} &=&
\det{
\left(
\begin{array}{cc}
j \omega C_1 &R_1^{-1} \\
Y_{41} &  j \omega C_2 
\end{array}
\right)
} \nonumber \\
&=&
-\omega^2 C_1C_2-R_1^{-1}Y_{41}.
\end{eqnarray}
The determinant is zero if and only if
\begin{equation}
 Y_{41}=-\omega^2 C_1 C_2  R_1.
\end{equation}
Thus, $Y_{41}$ must be a negative quantity. In the next section, we discuss how to implement a negative resistance standard before we return to a description of the complete four-arm bridge.

\section{Apparent Negative Resistance Standard}
\begin{figure}[htbp]
\begin{center}
\includegraphics[width=0.95\columnwidth]{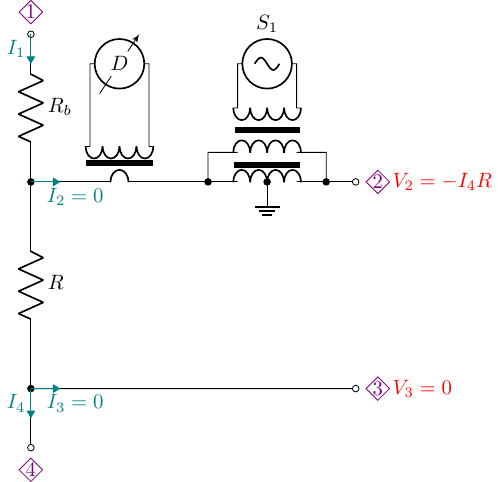}
\end{center}
\caption{Schematic of a negative resistance standard with an apparent value of $-R$. The current supplied to port 1 is adjusted such that $D$ is nulled.
\label{fig:negR}}
\end{figure}

Apparent negative impedances have been previously used to facilitate unloading when a 2TP bridge was used to compare 4TP impedances~\cite{Shields1974}. Here we adopt this proven concept to construct a 4TP negative resistance standard as illustrated in Fig.~\ref{fig:negR}, which will be shown in the next section to facilitate the comparison of resistance with capacitance. The idea is to convert a 4TP resistor of a nominal value of $R$ into a 4TP composite resistor with an apparent negative resistance, $-R$. The positive terminal of a one-to-one transformer with its center tap grounded is connected to the original high potential port of $R$ through a defining transformer. The negative terminal of the transformer then becomes the new high potential port 2. A desired output voltage of port 2 can be set by adjusting a waveform generator, $S_1$, that excites the one-to-one transformer through an isolation transformer. The current through the high current port 1 is supplied by another waveform generator (not shown in the figure), $S_b$, through an auxiliary resistor, $R_b\approx R$. Here, $S_1$ and $S_b$ are locked in frequency and phase. By adjusting $S_b$, the current through the defining transformer can be minimized. The apparent resistance of the composite standard is 
\begin{equation}
    R_\mathrm{app} = -R (1+\gamma),
\end{equation}
where $\gamma$ is a small complex number representing the ratio error of the one-to-one transformer. A reversal of the transformer provides self-calibration.

\section{Coaxial Circuit}

\begin{figure*}[ht!]
\begin{center}
\includegraphics[width=2\columnwidth]{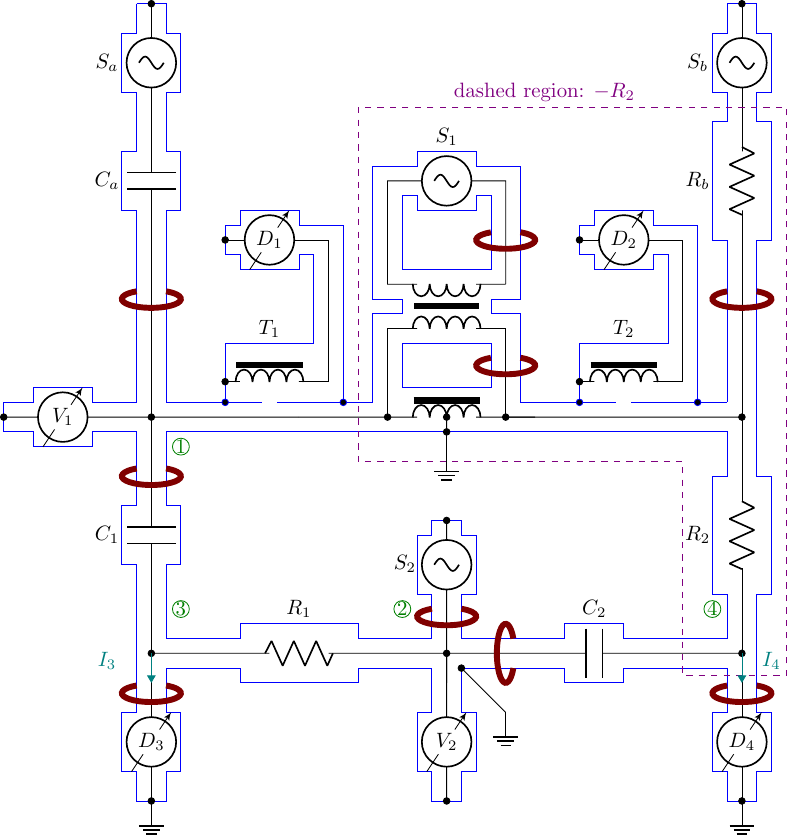}
\end{center}
\caption{Coaxial schematic of a four-arm bridge with unloading circuits.  The encircled round numbers correspond to the encircled nodes in Fig.~\ref{fig:simple4}. Detectors $D_3$ and $D_4$ are amplifiers with their inputs at virtual ground.
\label{fig:dia}}
\end{figure*}

The coaxial circuit of a digital four-arm bridge for the comparison of resistance with capacitance is shown in Fig.~\ref{fig:dia} with the outer conductor drawn in blue. The encircled numbers in the figure correspond to the nodes in Fig.~\ref{fig:simple4}. The left branch of the bridge consists of $Y_{31} = j\omega C_1$ and $Y_{32} = R_1^{-1}$, while the right branch consists of $Y_{41} = -R_2^{-1}$ and $Y_{42} = j\omega C_2$. The negative $R_2$  given within the dashed region is a composite negative resistor, as is shown in Fig.~\ref{fig:negR}. 
Both capacitors are 2TP standards. $C_1$ is defined between nodes 1 and 3, and the nominally equal $C_2$ between nodes 2 and 4. $R_1$ and $R_2$ are nominally equal 4TP standards. Mixing of 2TP capacitors with 4TP resistors in the same bridge has been demonstrated previously~\cite{Small2001,Thompson1968, Small1987}.

All $C$ and $R$ represent complex capacitances and resistances, respectively. $C_1$ is unloaded through an approximately equal capacitor, $C_a$, by adjusting its voltage source, $S_a$, such that the detector, $D_1$, of the defining transformer $T_1$ is nulled. The high port of $C_2$ is connected to the high potential port of $R_1$. The contact resistance between them is considered as a part of $C_2$ which may slightly alter its dissipation factor but contributes negligibly to its capacitance. The low terminal of $C_1$ is connected to the low current port of $R_1$ with the low potential port of $R_1$ kept at the virtual ground by a current amplifier (Femto DLPCA-200\footnote{Certain commercial equipment, instruments, or materials are identified in this paper to foster understanding. Such identification does not imply recommendation or endorsement by the National Institute of Standards and Technology, nor does it imply that the materials or equipment identified are necessarily the best available for the purpose.}). The connection between $R_2$ and $C_2$ is similarly arranged. 

The four waveform generators (Keysight 33500B), $S_1$, $S_2$, $S_a$, and $S_b$ are locked in frequency and phase through their external reference frequency inputs to a clock generator (Stanford Research Systems CG635), nominally set at \SI{10}{\mega\hertz}, which is then locked to a \SI{10}{\mega\hertz} GPS clock. The clock generator in the path of reference frequency provides the convenience of changing the bridge operating frequency by a small amount without disturbing the overall bridge balance, as has been done, for example, to acquire the data shown in Fig.~\ref{fig4}.  $S_1$ and $S_2$ are operated with nominal peak-to-peak amplitudes of \SI{10}{\volt}. 

A quadrature bridge for the comparison of resistance with capacitance has a distinctive advantage, compared to ratio bridges for the comparison of like impedances, in that the operating frequency can be continuously changed to match the impedance magnitude of the capacitor with that of the resistor~\cite{Delahaye1992}. The four-arm bridge is at its balance when $\det(\mathbf{Y})=0$. As detailed in Sec.~2, at this operating point, the virtual error current, $\Delta I$, is to first order independent of fluctuations in the supply voltages measured by $V_1$ and $V_2$, and the complex coefficients $\rho_1$ and  $\rho_2$ that are used to calculate the error current. Let $\omega_\circ$ denote the frequency at which the bridge is at balance. Then,
\begin{eqnarray}
\det{(
\mathbf{Y}_\circ)} &=&
\det{
\left(
\begin{array}{cc}
j \omega_\circ C_1 &R_1^{-1} \\
-R_2^{-1} & j \omega_\circ C_2
\end{array}
\right)
} \nonumber \\
&=&
-\omega_\circ^2C_1 C_2+R_1^{-1}R_2^{-1}=0,
\end{eqnarray}
which can be solve for $\omega_\circ$ to get
\begin{equation}
\omega_\circ =(R_1 R_2 C_1 C_2)^{-1/2}
\label{eq:det0}.
\end{equation}
A small built-in trim capacitor in parallel of $R_1$ can be adjusted to meet the balance condition for the imaginary component.
With $\mathbf{Y}_\circ$ set, the source balance, Eq.~(\ref{eq:sb}), gives the value of $\mathbf{V}_\circ$ as
\begin{equation}
\mathbf{V}_\circ \approx V_1
\left(
\begin{array}{c}
1\\
-j
\end{array}
\right).
\label{eq:V0}
\end{equation}
Similarly, the detector balance, Eq.~(\ref{eq:db}), gives the value of $\boldsymbol\rho_\circ$ according to
\begin{equation}
\boldsymbol\rho_\circ  \approx 
\left(
\begin{array}{cc}
-j&
1
\end{array}
\right).
\label{eq:rho0}
\end{equation}

Without loss in generality, we assume that the two impedances are identical,  i.e., $C_\circ$ and $R_\circ$. To balance the bridge, the frequency is changed until $\Delta I\approx 0$ which is achieved at  $\omega_\circ R_\circ C_\circ=1$. Unbalance in the bridge can occur in all three factors  $R$, $C$, or $\omega$. Let's assume, $C$ is unchanged and $R=R_\circ(1+\alpha+j\beta)$, where $\alpha$ and $\beta$ are small real numbers.
The new balance condition is, 
\begin{equation}
\omega R C_\circ =1 \longrightarrow \omega = \frac{1}{C_\circ R}\approx \omega_\circ (1-\alpha -j\beta).
\end{equation}
Similarly, a small fractional change in frequency corresponds to an identical fractional change in the real part of the resistor.

If a small change occurs after the bridge has been balanced and the operator does not choose to rebalance the bridge by changing its frequency, the error current gives the fractional change directly. Using $\boldsymbol\rho_\circ$ from Eq.~(\ref{eq:rho0}) and $\mathbf{V}_\circ$ from Eq.~(\ref{eq:V0}), one finds
\begin{equation}
\Delta I=\boldsymbol\rho_\circ\mathbf{Y} \mathbf{V}_\circ = \frac{2  V_1}{R_\circ} (\alpha + j \beta). 
\end{equation}
Multiplying $\Delta I$  with $R_\circ/(2 V_1) =1/(2 \omega_\circ  V_1 C_\circ)$, the small deviations $\alpha$ and $\beta$ are directly obtained from the error current,
\begin{equation}
\alpha + j \beta = \frac{ \Delta I}{2 \omega_\circ V_1 C_\circ }.
\label{eq:eps:alpha}
\end{equation}

In practice, $\alpha + j\beta$ is a null indicator that can be aligned and calibrated, like a lock-in detector, by a known displacement of the bridge balance such as changing $\omega$ slightly. The current detectors $D_3$ and $D_4$ are individually aligned to the natural fluctuations of $V_1/V_2$ through correlation analysis~\cite{Feige2022}.
The ratio $\rho_1/\rho_2$  is determined by analyzing the correlation between $I_3$ and $I_4$, which are digitized synchronously using a Keysight DAQM909A. 

The unloading voltage sources $S_a$ and $S_b$ only need to be adjusted once to null the detectors, $D_1$ and $D_2$, respectively. The stabilities of $S_a$ and $S_b$ are not critical which we have investigated experimentally. Turning $S_a$ and $S_b$ completely off only changes the bridge result on the order of \SI{1}{\micro \ohm / \ohm}.

The voltage ratio $V_1/V_2$ is dynamically adjusted by modifying only $S_2$, with $S_1$ fixed, using a simple digital proportional-integral feedback algorithm, to minimize $I_3$. The feedback and its performance are described in detail in~\cite{Feige2022}.
Although the peak-to-peak fluctuations of $V_1$ and $V_2$ remain on the order \SI{100}{\micro \volt / \volt}, their statistical means deviate less than \SI{1}{\micro \volt / \volt} from their ideal values required by the source balance.

The bridge balance signal is not sensitive to a small change of $\delta \mathbf{V}$ according to Eq.~(\ref{eq:epsfull}). This was verified by deliberately setting $S_2$ incorrectly either in amplitude or phase. For example, adding \SI{10}{\micro \volt / \volt} to $V_2$  did not result in a detectable change in $\alpha$ or $\beta$.

\section{Bridge Performance}

The bridge was used to compare two \SI{1}{\nano\farad} capacitors with two \SI{100}{\kilo\ohm} resistors at a frequency near \SI{1592}{\hertz}. One of the capacitors, $C_1$, consists of ten \SI{100}{\pico \farad} fused-silica capacitors in parallel, while the other, $C_2$, is a temperature-regulated \SI{1}{\nano\farad} ceramic capacitor. The capacitance difference between $C_1$ and $C_2$ is less than \SI{4}{\micro\farad/\farad}, and their stability is better than \SI{1}{\nano\farad/\farad} during the experiments. The two resistors are closely matched (within \SI{12}{\micro\ohm/\ohm}) Vishay metal-foil resistors that are hermetically sealed in oil-filled cans. Each resistor is individually housed in a thick aluminum box whose temperature is indirectly regulated by seating on top of a large aluminum block attached to a temperature-controlled breadboard (Thorlabs  PTC1). Although the specified stability of the breadboard is only \SI{0.1}{\celsius}, the temperature fluctuation of the resistor housings is about \SI{0.02}{\celsius}, estimated by attaching a calibrated thermistor probe.

\begin{figure}[htbp]
\begin{center}
\includegraphics[width=0.95\columnwidth]{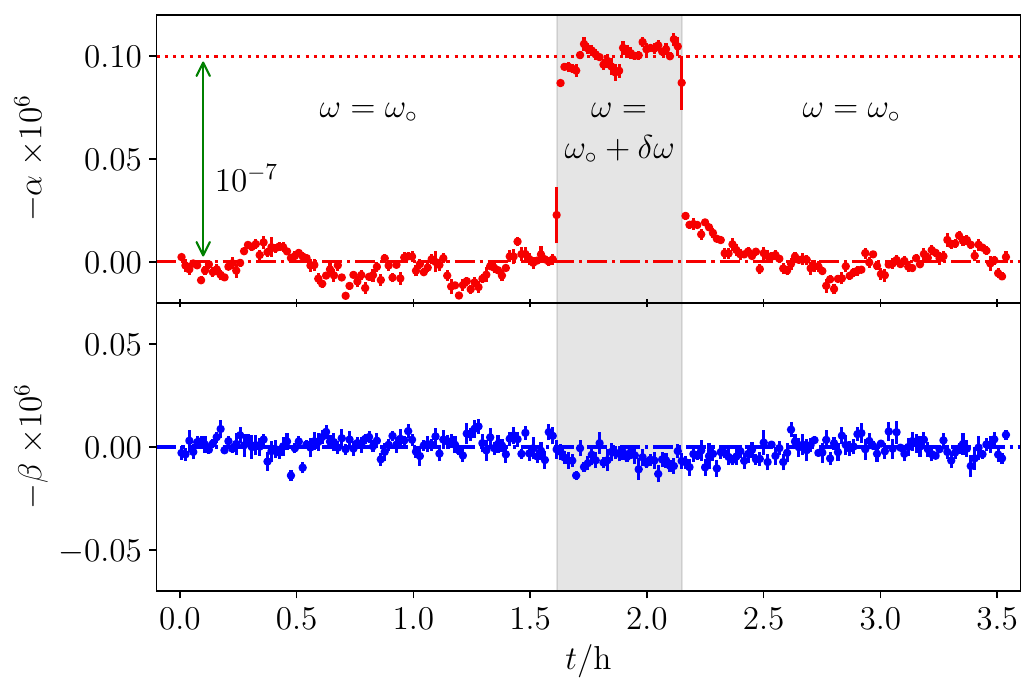}
\end{center}
\caption{Measurements of $\alpha$ (top panel) and $\beta$  (bottom panel) as a function of time. After about 1.5 hours into the data taking the bridge frequency was changed from $\omega_\circ$ to $\omega_\circ+\delta \omega$, with $\delta \omega/\omega_\circ= \SI{1e-7}{}$. 
\label{fig4}}
\end{figure}

\begin{figure}[htbp]
\begin{center}
\includegraphics[width=0.95\columnwidth]{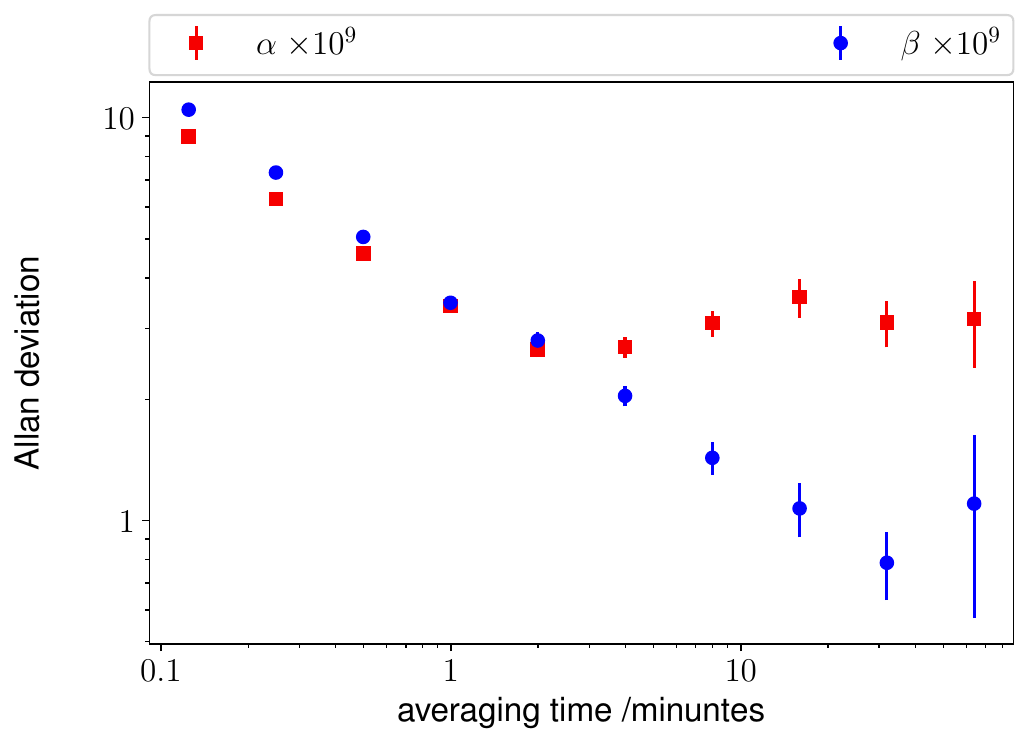}
\end{center}
\caption{Squares and circles give the Allan deviations of $\alpha$ and $\beta$ as a function of measurement time. Error bars are $1$-$\sigma$ standard deviation of the Allan deviation. 
\label{fig5}}
\end{figure}

Measured values of  $\alpha$ and $\beta$ are shown in Fig.~\ref{fig4} as a function of time, demonstrating the stability of the bridge. In the middle of the test run, we increased the clock generator frequency by \SI{1}{\hertz}, inducing a step change of \SI{1e-7}{} in $\alpha$ without disturbing $\beta$. The degree of orthogonality between $\alpha$ and $\beta$ was further tested by increasing $\beta$ as much as $\SI{1e-5}{}$ without noticing any discernible change in $\alpha$. We conclude that the quadrature component of the bridge does not need to be fully nulled since the primary focus is the determination of resistance in terms of frequency and capacitance. Shown in Fig.~\ref{fig5} are the typical Allan deviations of $\alpha$ and $\beta$ measured after balancing the bridge at $\omega_\circ = \SI{1591.9520}{\hertz}$ as a function of averaging time. The relative Allan deviation of both $\alpha$ and $\beta$ are about \SI{1e-8}{} for the first point, acquired in \SI{7}{\second}, and they initially decrease along a straight line in the log-log plot, with their slopes consistent with averaging over white noise. After one minute of averaging $\alpha$ drops to below \SI{3e-9}{}, then increases slightly in \SI{10}{\minute}. This increase is probably due to the imperfect temperature control of the resistors. Furthermore, $\beta$ reaches below \SI{1e-9}{}, suggesting that the bridge structure and the computer algorithm are very stable.

Table~\ref{tab1} summarizes the uncertainty budget for the measurement of $\alpha$. The contribution of the digitizer nonlinearity is negligible because $\alpha$ was nulled. A possible contribution of $\beta$ to $\alpha$ was estimated from the residual  $\beta$ and the determined degree of the orthogonality. The ratio error of the one-to-one transformer was determined by reversing the transformer.

A quadrature bridge for the RC comparison is frequency-dependent and has a unique problem due to the nonlinear mixing of harmonics at the detector. The theoretical framework for dealing with the resultant intermodulation offset can be found in the literature~\cite{Awan2011}. The digital bridge has an advantage because all relevant harmonics are simultaneously digitized with the fundamental signal. We followed the procedure described in~\cite{Small2001} to determine the intermodulation offset.

We used analog filters (Stanford Research Systems SIM965) between the current amplifiers and the digitizer so that the harmonic intermodulation was limited to the current amplifiers. By changing the gains of the two current amplifiers by a factor of 100, we noticed a change of \SI{2e-8} in $\alpha$, attributable to the intermodulation offset. To correct the offset, we first injected additional second and third harmonic signals, separately to each branch, to minimize the associated harmonic amplitude at the digitizer, and we then increased the injection amplitude by a factor of ten in order to measure the additional offset to provide the calibration factor. 

The choke errors were treated in a similar fashion, which has been thoroughly described in the literature~\cite{Awan2011,Thompson1968,Homan1968}. The combined standard uncertainty ($k = 1$) of the digital four-arm bridge is \SI{5E-9}{}, which compares favorably to the uncertainty of the conventional quad bridge at NIST reported previously~\cite{Jeffrey1998}.

\begin{table}
\caption{Uncertainty budget ($k = 1$) for $\alpha$.\label{tab1}}
\begin{tabular}{l|r}
Item & Std. Unc. $\times 10^9$\\
\hline
     Type A& 3   \\
     Ratio error of 1:1 transformer & 3\\
     Harmonic intermodulation offset& 1   \\
     Coaxial chokes & 1   \\
     Orthogonality error& 0.3\\
     Digitizer nonlinearity & $< 0.1$\\
    \hline
    Combined standard uncertainty & 5
\end{tabular}
\end{table}

The combined standard uncertainty does not include transport uncertainties of the standards which have to be included if the determined impedance ratio were to be compared with another method. In particular, the transport stability of the \SI{100}{\kilo\ohm} Vishay resistors has not been fully determined. We intend to use the impedance ratio determined by the digital four-arm bridge as a reference to improve a custom detection system for one-to-one complex voltage ratio which is the core technology and the limiting factor of the two-arm digital bridges described previously ~\cite{Feige2022,Wang2023}. For this limited application, the standards can stay unperturbed from one bridge configuration to another.

\section{Conclusion}

A digital four-arm  bridge for the comparison of resistance with capacitance has been demonstrated. The digital four-arm  bridge retains the key feature of the classical quad bridge with the source balance, the detector balance, and the main balance, where the required precision of the source voltages is only of the order of the square root of the ultimate bridge precision. For the comparison of a \SI{100}{\kilo\ohm} resistor with a \SI{1}{\nano\farad} capacitor near \SI{1592}{\hertz}, the combined standard uncertainty ($k = 1$) is \SI{5e-9}{}, which is comparable to the uncertainty of the conventional quad bridge at NIST reported previously~\cite{Jeffrey1998}. The impedance ratio determined by the digital four-arm bridge will provide an independent check for the two-arm digital bridges that rely on the digitizers to provide the one-to-one ratio reference~\cite{Feige2022,Wang2023}.

\section*{References}


\providecommand{\newblock}{}

\end{document}